\documentclass[
reprint, 
tightenlines, 
superscriptaddress, 
showpacs,%preprintnumbers,
nobibnotes,
 amsmath,amssymb,
 aps,
prl,
floatfix,
]{revtex4-1}

\usepackage{latexsym} 
\usepackage{enumerate}
\usepackage{bm}% bold math
\usepackage{graphicx}% Include figure files
\usepackage[usenames]{color} 
\usepackage{dcolumn}% Align table columns on decimal point
\begin{document}

%\preprint{APS/123-QED}

\title{Precision Measurements Using Squeezed Spin States via Two-axis Counter-twisting Interactions}
% Force line breaks with \\

\author{Emi Yukawa} 
\affiliation{National Institute for Informatics, 2-1-2 Hitotsubashi, Chiyoda-ku, Tokyo 101-8430, Japan}
\author{Gerard Milburn} 
\affiliation{Centre for Engineered Quantum Systems, School of Mathematics and Physics, 
The University of Queensland, St Lucia, QLD 4072, Australia}
\author{Masahito  Ueda} 
\affiliation{Department of Physics, University of Tokyo, 7-3-1 Hongo, Bunkyo-ku, Tokyo 113-0033, Japan}
\author{Kae Nemoto} 
\affiliation{National Institute for Informatics, 2-1-2 Hitotsubashi, Chiyoda-ku, Tokyo 101-8430, Japan}
%\affiliation{%


\date{\today}% It is always \today, today,

\begin{abstract} 
We show that the two-axis counter twisting interaction squeezes a coherent spin state into three states of 
interest in quantum information, namely, the twin-Fock state, the equally-weighted superposition state, and the state 
that achieves the Heisenberg limit of optimal sensitivity defined by the Cram\'er-Rao inequality in addition to the 
well-known Heisenberg-limited state of spin fluctuations. 
%\begin{description}
% Secondary publications and information retrieval purposes.
%03.75.Kk, 03.75.Mn, 05.30.Jp
% You may use the \texttt{description} environment to structure your abstract;
% use the optional argument of the \verb+\item+ command to give the category of each item. 
%\end{description}
\end{abstract}

\maketitle

%\tableofcontents

%\section{\label{sec:1}Introduction}  
Squeezed states have been intensively investigated originally in optics and then extended to various bosonic and spin systems.  
A defining feature of squeezing is to enhance the quantum nature such as reduced quantum noise and entanglement, which 
form the basis of their applications, for instance, high precision measurements~\cite{Aasi,Bowen}. 
Although entanglement is not always the key in high precision measurement~\cite{Tilma}, some of these implementations are 
expected to surpass the standard quantum limit. 

There are other quantum states proposed for high precision measurement, such as a superposition state of coherent states~\cite{Stoler}, squeezed spin states (SSSs)~\cite{Ueda}, and other spin ensemble states~\cite{Burnett,Heinzen,Zoller,Lukin1,Wineland,Maccone,Klempt}. 
These states may also achieve sensitivity beyond the standard quantum limit. 
Amongst them the advantage of spin squeezing is its feasible implementation of the state~\cite{Polzik,Bigelow,Mabuchi,Oberthaler1,Vuletic,Oberthaler2,Treutlein,Thompson,Chapman,Takahashi,Lukin2}.  
For instance, spin squeezing by the one-axis twisting has been experimentally realized in cold-atomic systems~\cite{Bigelow,Mabuchi,Oberthaler1,Vuletic,Oberthaler2,Treutlein,Thompson,Takahashi} and has been proposed in nitrogen-vacancy-spin ensembles~\cite{Lukin2}. 
Furthermore spin fluctuations below the standard quantum limit have been observed~\cite{Oberthaler2,Treutlein}. 
Meanwhile, spin squeezing by the two-axis counter twisting method~\cite{Ueda}, has not been realized; 
however, there are some experimental proposals~\cite{You1,Lukin1,Leung,You2,You3}. 

The minimum quantum fluctuations of the SSSs have been deeply investigated; however, the sensitivity of measurements utilizing the 
SSSs, which should also be discussed in terms of the estimation theory~\cite{Smerzi,Geremia,Nori}, has not been fully analyzed. 
The precision of a measurement may be defined by the standard deviation of a measurement-and-estimation process, which satisfies the Cram\'er-Rao inequality~\cite{Rao,Cramer}.  
Similarly to the Heisenberg limit for quantum spin fluctuations, we can define the Heisenberg limit of the minimum standard deviation using the Cram\'er-Rao inequality. 
For the total spin size of $J$ of spin-$1/2$ particles, a coherent spin state (CSS) gives the minimum standard deviation to be proportional to $J^{-1/2}$, and the one-axis twisting brings the state to achieve the Heisenberg limit, i.e. $J^{-1}$~\cite{Smerzi}. 

In this paper, we investigate time evolution of SSSs in a spin-1/2 ensemble through two-axis counter-twisting interactions~\cite{Ueda} and the optimal sensitivity in high precision measurement. 
Similarly to the one-axis twisting case~\cite{Smerzi}, we can expect that the SSS will change the sensitivity though time evolution. 
We will numerically show that depending on the evolution time of the two-axis counter twisting interaction, we can generate a SSS that has high fidelity to the 
equally-weighted superposition state (EWSS) or the twin-Fock state~\cite{Burnett,Klempt}, 
which are often proposed for high precision measurements with the Heisenberg-limited sensitivity and are not always easy to be implemented. 
We will also numerically obtain the sensitivities for these SSSs and the SSS optimizing the sensitivity with respect to the evolution time and 
compare them with the EWSS, the twin-Fock state, and the cat state. 

%\section{\label{sec:2}Time evolution of a coherent spin state under the two-axis counter twisting interactions} 
We consider a SSS generated from a CSS via the two-axis counter twisting interaction. 
Here, we assume an ensemble of $N$ spin-1/2 particles as a system. 
With two real parameters $\alpha$ and $\beta$, a CSS is given by 
$|{\Psi}_{\mathrm{CSS}} (\alpha ,\beta ) \rangle = \bigotimes_{i=0}^N |\psi (\alpha ,\beta ) {\rangle}_i$, 
where  $\ |\psi (\alpha ,\beta ) {\rangle}_i = \cos {\frac{\beta}{2}} \ |\uparrow {\rangle}_i + e^{i\alpha} \sin {\frac{\beta}{2}} \ |\downarrow {\rangle}_i$.  
The suffix $i$ denotes the $i$-th $1/2$ spin and $|\uparrow {\rangle}_i$ and $|\downarrow {\rangle}_i$ are the eigenstates of the Pauli matrix ${\hat{\sigma}}_z^{(i)}$ for the eigenvalues of  $\pm 1$, respectively.
We introduce the collective spin operator $\hat{\bm{J}} \equiv \sum_{i=0}^N \frac{1}{2}{\hat{\bm{\sigma}}}^{(i)}$ and expand the CSS 
in terms of the eigenstates of ${\hat{J}}_z$, yielding  
\begin{equation} 
\begin{split}
	|{\Psi}_{\mathrm{CSS}} (\alpha ,\beta ) \rangle = \sum_{M=-J}^J & \sqrt{\binom{2J}{J-M}} \ e^{i(J-M)\alpha } \\ 
	\times & {\cos}^{J+M} \frac{\beta}{2} \ {\sin}^{J-M} \frac{\beta}{2} \ |J,M \rangle ,\label{eq:CSS2}
\end{split} 
\end{equation} 
where $J=N/2$ and $|J,M\rangle$ represents the eigenstate of ${\hat{J}}_z$ with the eigenvalue of $M$. 
We set $\alpha$ and $\beta$ to be zero such that the initial CSS is fixed to $|J,J\rangle$. 
Then, the two-axis counter twisting Hamiltonian can be expressed in terms of the collective spin operators as 
\begin{equation} 
	{\hat{H}}_{\mathrm{TAT}} = \frac{\hbar \chi}{2i} \left ( e^{-2i\gamma} {\hat{J}}_{+}^{2} - e^{2i\gamma} {\hat{J}}_{-}^{2} \right ), 
	\label{eq:TAT} 
\end{equation} 
where $\chi$ is the strength of interaction, ${\hat{J}}_{\pm}$ denotes the spin-$J$ ladder operators, and $\gamma$ determines the 
orientation of the spin squeezing.
For the sake of simplicity, we choose $\chi =1$ and $\gamma =0$, setting the squeezing (anti-squeezing) axis to be $J_y$($J_z$).
We begin with the initial coherent state $|J,J\rangle$, let it evolve for a certain time $\tau$, and then rotate it along the $y$-axis by $\pi/2$, hence the resulting state is 
\begin{equation} 
	|{\Psi}_{\mathrm{SSS}} (\tau) {\rangle}_x = \exp {\left [ -i\frac{\pi}{2} {\hat{J}}_y \right ]}\exp {[{\hat{H}}_{\mathrm{TAT}} \tau /i \hbar ]} 
	|J,J \rangle . \label{eq:SSSx}
\end{equation} 

%\section{\label{sec:3}Comparison with the equally-weighted superposition state and the twin-Fock state}  
First, we evaluate the SSS in Eq.~(\ref{eq:SSSx}) at certain evolution time $\tau$, when the SSS has the optimal fidelity 
to the EWSS or the twin-Fock state~\cite{Burnett,Klempt}. 
Here, the fidelity of the SSS in Eq.~(\ref{eq:SSSx}) to the state $|{\Psi}_X \rangle$ is given by 
$F_X (\tau ) = |\langle {\Psi}_{X} | {\Psi}_{\mathrm{SSS}} (\tau ) {\rangle}_x |^2$ as a function of the evolution time $\tau$ 
in Eq.~(\ref{eq:SSSx}). 
The EWSS and the twin-Fock state are given by 
\begin{equation} 
	|{\Psi}_{\mathrm{EWSS}} \rangle \equiv \frac{1}{\sqrt{2J+1}} \sum_{M=-J}^J |J,M \rangle , \label{eq:EWSS} 
\end{equation} 
and 
\begin{equation} 
	|{\Psi}_{\mathrm{TFS}} \rangle \equiv \exp {\left [ - i\frac{\pi}{2} {\hat{J}}_x \right ]} \ |J,0 \rangle , \label{eq:TFS} 
\end{equation} 
respectively. 
\begin{figure}[t] 
\begin{center} 
\includegraphics[width=7.5cm,clip]{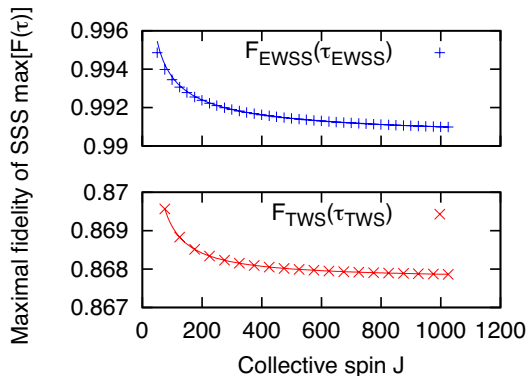}
\caption{Maximal fidelity of the SSS to the EWSS and that to the twin-Fock state as functions of the 
collective spin $J$. 
The blue solid curve is given by Eq.~(\ref{eq:FmaxtoEWSS}) and well fits to the region of $J \geq 400$, 
while the red solid curve is given by Eq.~(\ref{eq:FmaxtoTFS}) and shows excellent agreement with the plot points.} 
\label{fig:FEWSSandFTFS} 
\end{center} 
\end{figure} 
The fidelity functions $F_{\mathrm{EWSS}} (\tau )$ and $F_{\mathrm{TFS}} (\tau )$ are maximized at certain evolution 
times ${\tau}_{\mathrm{EWSS}}$ and ${\tau}_{\mathrm{TFS}}$ for a fixed collective spin $J$. 
We refer the SSSs at evolution time of ${\tau}_{\mathrm{EWSS}}$ and ${\tau}_{\mathrm{TFS}}$ as the SSSs optimized to 
the EWSS and the TFS, respectively. 
We numerically obtain $F_{\mathrm{EWSS}} ({\tau}_{\mathrm{EWSS}} )$ and $F_{\mathrm{TFS}} ({\tau}_{\mathrm{TFS}} )$ and 
plot them in Figs.~\ref{fig:FEWSSandFTFS}. 
The maximal fidelity $F_{\mathrm{EWSS}} ({\tau}_{\mathrm{EWSS}} )$ to the EWSS monotonically decreases with respect to $J$, which 
can be well fitted for $J \geq 400$ to 
\begin{equation} 
	F_{\mathrm{EWSS}} ({\tau}_{\mathrm{EWSS}}) = {\left (\frac{0.0298\pm0.0001}{J^{0.621\pm 0.001}} 
	+ 0.995 \right )}^2. \label{eq:FmaxtoEWSS}
\end{equation} 
Here, the numerical results throughout the paper contain numerical errors less than the order of the last digit, otherwise stated.
In the large-$J$ limit, $F_{\mathrm{EWSS}} ({\tau}_{\mathrm{EWSS}})$ in Eq.~(\ref{eq:FmaxtoEWSS}) converges to $\sim 0.990$, 
which is interesting as usually the EWSS is not easy to physically implement. 
Meanwhile, $F_{\mathrm{TFS}} ({\tau}_{\mathrm{TFS}} )$ monotonically decreases with 
respect to $J$ similarly to the EWSS case as shown in the lower panel of Figs.~\ref{fig:FEWSSandFTFS}, 
which can be well fitted to 
\begin{equation}
	F_{\mathrm{TFS}} ({\tau}_{\mathrm{TFS}}) = \left (\frac{0.0743}{J^{1.00}} + 0.932 \right )^2. \label{eq:FmaxtoTFS}
\end{equation} 
Equation~(\ref{eq:FmaxtoTFS}) converges to $\sim 0.868$ in the large $J$ limit. 
\begin{figure*}[t] 
\begin{center} 
\begin{tabular}{c}  

\begin{minipage}{0.24\hsize} 
\begin{center} 
\begin{flushleft} (a) \end{flushleft} 
\includegraphics[width=3.5cm,clip]{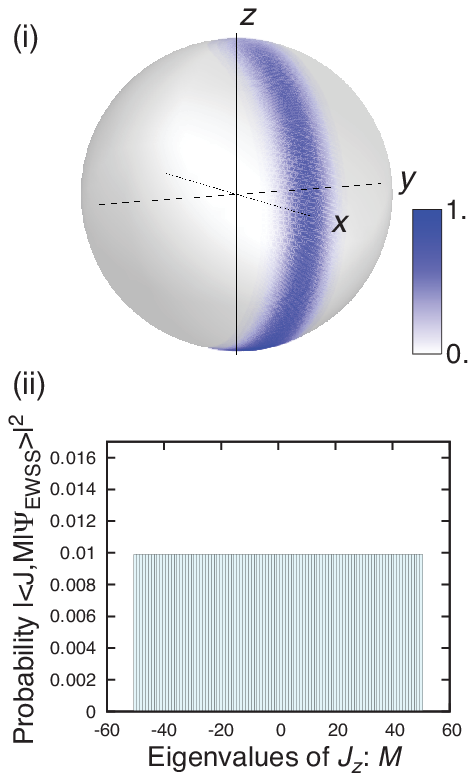}
\end{center}  
\end{minipage}  

\begin{minipage}{0.24\hsize} 
\begin{center}  
\begin{flushleft} (b) \end{flushleft} 
\includegraphics[width=3.5cm,clip]{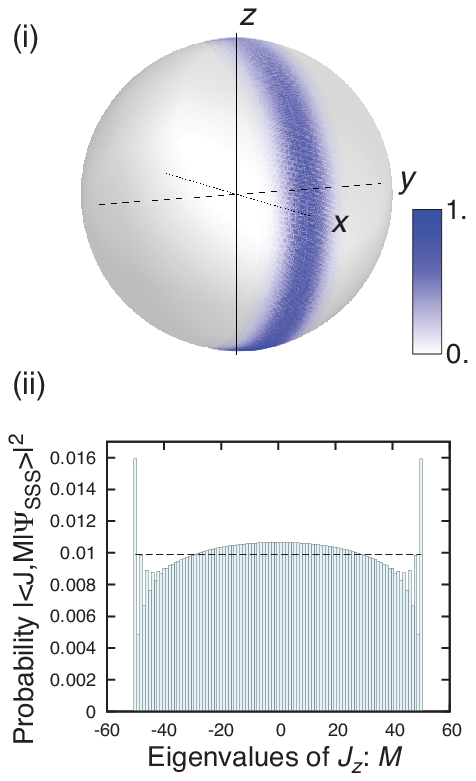}
\end{center}
\end{minipage}  

\begin{minipage}{0.24\hsize} 
\begin{center} 
\begin{flushleft} (c) \end{flushleft} 
\includegraphics[width=3.5cm,clip]{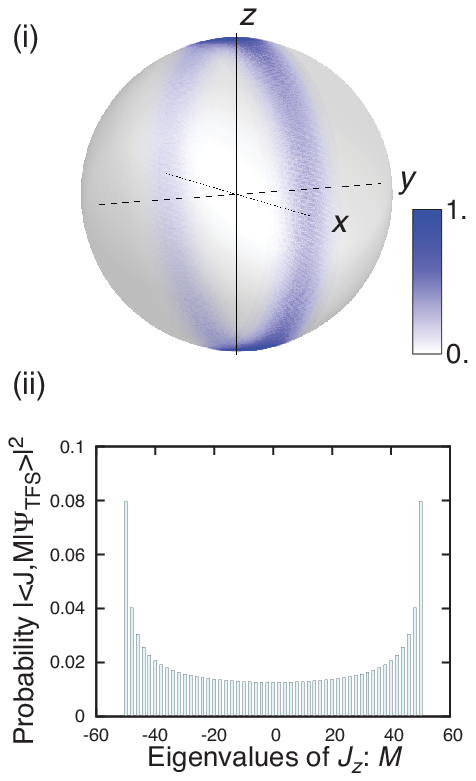}
\end{center}  
\end{minipage}  

\begin{minipage}{0.24\hsize} 
\begin{center} 
\begin{flushleft} (d) \end{flushleft} 
\includegraphics[width=3.5cm,clip]{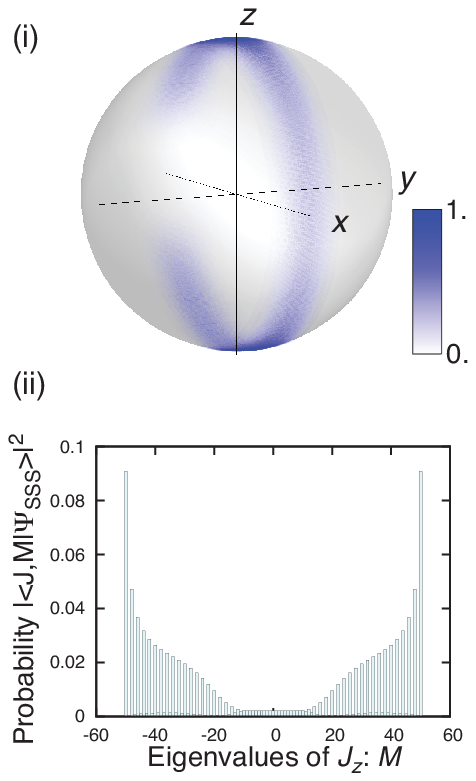}
\end{center}
\end{minipage}  

\end{tabular} 
\caption{QPD functions and probability distributions with respect to the eigenvalues of ${\hat{J}}_z$ 
for (a) the EWSS, (b) the SSS that maximizes the fidelity to the EWSS, (c) the twin-Fock state, and (d) the SSS that maximizes the 
fidelity to the twin-Fock state in the case of $J=50$. 
The QPDs are represented by the blue color density on the spheres in the upper panel (i). 
The probability distributions are shown in the lower panel (ii). 
The dashed line in (b)-(ii) indicates the probability distribution of the EWSS, that is, $(2J+1)^{-1}$. 
}
\label{fig:EWSSandTFSandSSS} 
\end{center} 
\end{figure*} 
To discuss the deviations of $F_{\mathrm{EWSS}} ({\tau}_{\mathrm{EWSS}})$ and $F_{\mathrm{TFS}} ({\tau}_{\mathrm{TFS}})$ 
from $F=1$, we plot the quasi-probability distribution (QPD) and the probability distributions for the SSSs optimized to the EWSS 
and the twin-Fock state for $J=50$ in Figs.~\ref{fig:EWSSandTFSandSSS}. 
Here, the QPD function and the probability distribution of the state $|{\Psi}_{X} \rangle$ are respectively defined as  
${\tilde{P}}_{X} (\varphi ,\theta ) = | \langle {\Psi}_{\mathrm{CSS}} (\varphi ,\theta ) |{\Psi}_{X} \rangle |^2$ and 
${P}_{X} (M) = | \langle J,M |{\Psi}_{X} \rangle |^2$, where $\varphi$ and $\theta$ in the QPD are the azimuth 
and polar angles of the sphere with the radius of $1$. 
The probability distribution of $|{\Psi}_{\mathrm{SSS}} ({\tau}_{\mathrm{EWSS}}) {\rangle}_x$ in Fig.~\ref{fig:EWSSandTFSandSSS} (b) (ii) oscillates around $|M| \sim J$, unlike 
that of the EWSS in Fig.~\ref{fig:EWSSandTFSandSSS} (a)-(ii), which may cause diminution of $F_{\mathrm{EWSS}} ({\tau}_{\mathrm{EWSS}})$. 
On the other hand, both of the QPD and the probability distribution of $|{\Psi}_{\mathrm{SSS}} ({\tau}_{\mathrm{TFS}}) {\rangle}_x$ are distinctive to that of the twin-Fock state as 
shown in Figs.~\ref{fig:EWSSandTFSandSSS} (c) and (d): the QPD of $|{\Psi}_{\mathrm{SSS}} ({\tau}_{\mathrm{TFS}}) {\rangle}_x$ in Fig.~\ref{fig:EWSSandTFSandSSS} 
(d) (i) shows a gap at $(\varphi ,\theta ) \sim (\pi ,\pi /2)$ and the probability distribution in Fig.~\ref{fig:EWSSandTFSandSSS} (d) (ii) has a dip structure around $M=0$, which 
contributes to the degradation of the fidelity for a large $J$. 

%\section{\label{sec:4}Precision of quantum measurement and estimation processes} 
Then, we evaluate the SSS for high precision measurements using estimation theoretic tools, namely, the Cram\'er-Rao inequality. 
The Cram\'er-Rao inequality gives the lower limit of the standard deviation~~\cite{Rao,Cramer,Caves,Milburn}. 
Let us assume that we perform a positive operator-valued measure (POVM) on a value $X$, repeating it $N_{\mathrm{msr}}$ times to 
estimate $X$ from $N_{\mathrm{msr}}$ outcomes. 
The deviation of $X$'s estimator $X_{\mathrm{est}}$ is defined as 
$\delta X \equiv X_{\mathrm{est}}/|d {\langle \! \langle X_{\mathrm{est}} \rangle \! \rangle}_X/dX| - X$, 
where ${\langle \! \langle \rangle \! \rangle}_X$ represents the expectation value of $N_{\mathrm{msr}}$-times measurements. 
The precision of the measurement, i.e., the standard deviation of the estimator $X_{\mathrm{est}}$, is given by 
${\sigma}_X \equiv ( N_{\mathrm{msr}} \langle \! \langle {(\delta X)}^2 {\rangle \! \rangle}_X)^{1/2}$. 
The standard deviation ${\sigma}_X $ satisfies the Cram\'er-Rao inequality, that is, 
\begin{equation}
	{\sigma}_X \geq \frac{1}{\sqrt{{\mathcal{I}}_X}}.  \label{eq:CRineq1}
\end{equation}
Here ${\mathcal{I}}_X$ is the quantum Fisher information~\cite{Caves,Milburn}, which has the upper bound determined by the input state interacting with $X$. 
Since we are interested in the fundamental properties of high precision measurements, we assume the input state to be pure 
${\hat{\rho}}_{\mathrm{inp}} = |{\Psi}_{\mathrm{inp}} \rangle \langle {\Psi}_{\mathrm{inp}} |$~\cite{Milburn,Smerzi,Geremia}. 
In this case, the quantum Fisher information satisfies 
\begin{equation}
	{\mathcal{I}}_X \leq 4 {\langle {\hat{\rho}}_{\mathrm{inp}}^{\prime 2} \rangle}_{\mathrm{inp}}, \label{eq:Fisherineq}
\end{equation} 
where the operator ${\hat{\rho}}_{\mathrm{inp}}^{\prime}$ is the $X$-derivative of ${\hat{\rho}}_{\mathrm{inp}}$ and 
the expectation value $\langle \hat{O} {\rangle}_{\mathrm{inp}} \equiv \mathrm{Tr} \ [{\hat{\rho}}_{\mathrm{inp}} \hat{O}]$.  
Combining inEqs.~(\ref{eq:CRineq1}) and (\ref{eq:Fisherineq}), we obtain the inequality satisfied by the standard deviation 
\begin{equation} 
	{\sigma}_X \geq \frac{1}{2 {\langle {\hat{\rho}}_{\mathrm{inp}}^{\prime 2} \rangle}_{\mathrm{inp}}^{1/2}}. \label{eq:CRineq2}
\end{equation} 
We use inEq.~(\ref{eq:CRineq2}) to obtain the optimal sensitivity for the estimation of the magnetic field $\bm{B}$ along the 
$z$-axis~\cite{Geremia,Nori}. 
The system evolves under the Hamiltonian, ${\hat{H}}_{\bm{B}} = - \hbar {\gamma}_s B {\hat{J}}_z$,  
where ${\gamma}_s$ denotes the gyromagnetic ratio and $B$ is the magnitude of $\bm{B}$. 
To estimate $B$, a state ${\hat{\rho}}_{\mathrm{inp}} (0)$ is prepared at the initial time, and the state after a certain time 
$t$ under ${\hat{H}}_{\bm{B}}$, ${\hat{\rho}}_{\mathrm{inp}} (t)$, is used as the input state of the parameter estimation. 
Since the upper bound of the Fisher information is given by Eq.~(\ref{eq:Fisherineq}), we obtain 
${\mathcal{I}}_B (t) \leq 4 {({\gamma}_s t)}^2 \langle {(\Delta J_z)}^2 {\rangle}_{\mathrm{inp}}$ by substituting 
${\hat{\rho}}_{\mathrm{inp}} (t)$ into the inequality~(\ref{eq:Fisherineq}). 
Here the quantum fluctuations in an observable $\hat{O}$ is defined as 
$\langle (\Delta O)^2 \rangle \equiv \langle {\hat{O}}^2 \rangle - \langle \hat{O} {\rangle}^2$. 
The Cram\'er-Rao inequality in Eq.~(\ref{eq:CRineq2}) is now given by 
\begin{equation} 
	{\sigma}_B \geq \frac{1}{2{\gamma}_s t \langle {(\Delta J_z)}^2 {\rangle}_{\mathrm{inp}}^{1/2}},   
	\label{eq:CRineq-B}
\end{equation}  
implying that the measurement precision is determined by $\langle {(\Delta J_z)}^2 {\rangle}_{\mathrm{inp}}^{1/2}$ and 
the Heisenberg-limited sensitivity can be achieved when $\langle {(\Delta J_z)}^2 {\rangle}_{\mathrm{inp}}^{1/2}$ is 
linear to $J$; hence we analyze the scaling law of $\langle {(\Delta J_z)}^2 {\rangle}_{\mathrm{inp}}^{1/2}$ in stead of 
inEq.~(\ref{eq:CRineq-B}) to discuss the sensitivity. 

\begin{figure*}[t] 
\begin{center} 
\begin{tabular}{c}  

\begin{minipage}{0.48\hsize} 
\begin{center} 
\begin{flushleft} (a) \end{flushleft} 
\includegraphics[width=7.5cm,clip]{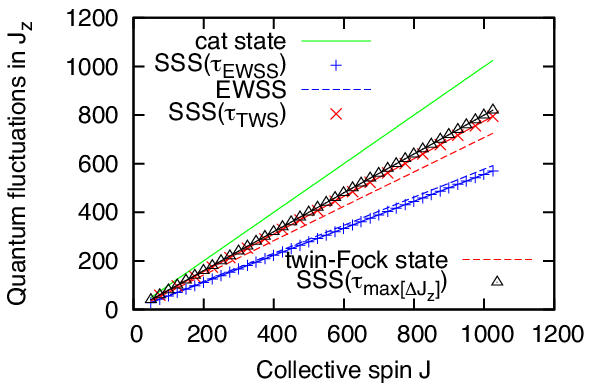}
\end{center}  
\end{minipage}  

\begin{minipage}{0.48\hsize} 
\begin{center}  
\begin{flushleft} (b) \end{flushleft} 
\includegraphics[width=7cm,clip]{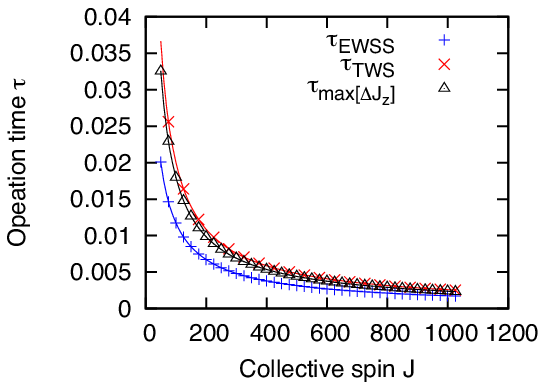}
\end{center}
\end{minipage}  

\end{tabular} 
\caption{Quantum fluctuations in ${\hat{J}}_z$ for the SSSs and their corresponding evolution 
time as functions of the collective spin $J$. 
(a) $\langle (\Delta {\hat{J}}_z )^2 {\rangle}_{\mathrm{SSS}}^{1/2}$ for the SSSs optimized to the EWSS and the twin-Fock state, 
and the maximal quantum fluctuations in ${\hat{J}}_z$ being compared with those 
for the cat state, the EWSS, and the twin-Fock state. 
Quantum fluctuations $\langle (\Delta {\hat{J}}_z )^2 {\rangle}_{\mathrm{SSS}}^{1/2} ({\tau}_{\mathrm{EWSS}})$ is well fitted 
to Eq.~(\ref{eq:DzSSSEWSS}) for $J \geq 400$, and 
$\langle (\Delta {\hat{J}}_z )^2 {\rangle}_{\mathrm{SSS}}^{1/2} ({\tau}_{\mathrm{TFS}})$ and the maximal 
$\langle (\Delta {\hat{J}}_z )^2 {\rangle}_{\mathrm{SSS}}^{1/2}$ are well fitted for all $J$ to Eqs.~(\ref{eq:DzSSSTFS}) and 
(\ref{eq:DzmaxSSS}), respectively. 
(b) Squeezing-evolution time ${\tau}_{\mathrm{EWSS}}$, ${\tau}_{\mathrm{TFS}}$, and ${\tau}_{\Delta J_z}$. 
The result are well fitted to Eqs.~(\ref{eq:FmaxtoEWSStime}), (\ref{eq:FmaxtoTFStime}), and (\ref{eq:DzmaxSSStime}), which are 
plotted by the solid curves for all $J$'s. 
} 
\label{fig:tandmaxDz} 
\end{center} 
\end{figure*}  
The quantum fluctuations in ${\hat{J}}_z$ are numerically calculated for 
the SSSs optimized to the EWSS and the twin-Fock state, and the SSS maximizing 
$\langle {(\Delta J_z)}^2 {\rangle}_{\mathrm{inp}}^{1/2}$ with respect to $\tau$ and they are plotted in Fig.~\ref{fig:tandmaxDz} (a). 
For the SSS optimized to the EWSS, $\langle {(\Delta J_z)}^2 {\rangle}_{\mathrm{SSS}}^{1/2} ({\tau}_{\mathrm{EWSS}})$ 
is well fitted for $J \geq 300$ to 
\begin{equation} 
	\langle {(\Delta J_z)}^2 {\rangle}_{\mathrm{SSS}}^{1/2} ({\tau}_{\mathrm{EWSS}}) = 0.557 
	(J+1.03)^{1.00},  \label{eq:DzSSSEWSS} 
\end{equation} 
which can achieve the Heisenberg-limited sensitivity in Eq.~(\ref{eq:CRineq-B}). 
The linear coefficient in Eq.~(\ref{eq:DzSSSEWSS}) is close to that for the EWSS as shown in Fig.~\ref{fig:tandmaxDz} (a), 
since $\langle {(\Delta J_z)}^2 {\rangle}_{\mathrm{EWSS}}^{1/2} = \sqrt{J(J+1)/3} \simeq 0.577 J$ in the large-$J$ 
limit. 
The difference in the linear coefficients for the EWSS and $|{\Psi}_{\mathrm{SSS}} ({\tau}_{\mathrm{EWSS}}) \rangle$ 
in Eq.~(\ref{eq:DzSSSEWSS}) is as small as $3.59\%$, which indicates the spin squeezing through 
the two-axis counter twisting can be used as a good approximation of the EWSS for high precision measurements. 
Similarly, $\langle {(\Delta J_z)}^2 {\rangle}_{\mathrm{SSS}}^{1/2} ({\tau}_{\mathrm{TFS}})$ is well fitted to the function 
that shows the Heisenberg-limit scaling of the sensitivity: 
\begin{equation} 
	\langle {(\Delta J_z)}^2 {\rangle}_{\mathrm{SSS}}^{1/2} ({\tau}_{\mathrm{TFS}}) 
	= 0.775{(J+0.494)}^{1.00}, \label{eq:DzSSSTFS} 
\end{equation} 
which is smaller than the that for the twin-Fock state by a factor of $0.912$, since 
$\langle {(\Delta J_z)}^2 {\rangle}_{\mathrm{TFS}}^{1/2} \simeq \sqrt{J(J+1)/2} \simeq 0.707 J$ in the large-$J$ limit~\cite{Lukin1,Nori} 
as shown in Fig.~\ref{fig:tandmaxDz} (a). 
The quantum fluctuations $\langle {(\Delta J_z)}^2 {\rangle}_{\mathrm{SSS}}^{1/2}$ maximized with respect to $\tau$ is also well fitted 
to the Heisenberg-limit scaling of the sensitivity, that is, 
\begin{equation} 
	\langle {(\Delta J_z)}^2 {\rangle}_{\mathrm{SSS}}^{1/2} ({\tau}_{\mathrm{max}\Delta J_z}) = 0.799 {(J+0.453)}^{1.00}, 
	\label{eq:DzmaxSSS} 
\end{equation} 
where ${\tau}_{\mathrm{max}\Delta J_z}$ represents the corresponding evolution time. 
%\begin{table}[b]
%\caption{Comparison for coefficients of $J$ in $\langle (\Delta J_z )^2 {\rangle}^{1/2}$ for the EWSS, the SSS at 
%${\tau}_{\mathrm{EWSS}}$, the twin-Fock state, and the SSS at ${\tau}_{\mathrm{TFS}}$. }
%\begin{ruledtabular}
%\begin{tabular}{cc} 
% & $\langle {(\Delta J_z)}^2 {\rangle}^{1/2}$ in the large $J$ limit \\ \hline 
%GHZ & $J$ \\ \hline
%SSS(${\tau}_{\mathrm{max}[\Delta J_z]}$) & $0.799J$ \\ \hline
%EWSS & $0.577J$ \\
%SSS(${\tau}_{\mathrm{EWSS}}$) & $0.557J$ \\ \hline
%twin-Fock state & $0.707J$ \\
%SSS(${\tau}_{\mathrm{TFS}}$) & $0.775J$  
%\end{tabular}
%\end{ruledtabular}
%\label{tbl:1}
%\end{table}
The Cram\'er-Rao inequality in Eq.~(\ref{eq:CRineq-B}) gives the best precision when the state is the cat state (or the GHZ state), namely, the superposition state of the highest and lowest weight states, i.e. 
$|{\Psi}_{\mathrm{CAT}} \rangle = (|J,J\rangle + |J,-J \rangle )/\sqrt{2}$.
and the quantum fluctuations in ${\hat{J}}_z$ is given by $\langle {(\Delta J_z)}^2 {\rangle}_{\mathrm{CAT}}^{1/2} = J$~\cite{Nori}. 
It is clear that the sensitivity using the SSSs cannot reach the best sensitivity that the optimal superposition state gives for the same $J$. 
However, the sensitivity by the SSSs can achieve higher sensitivities than ones achievable by the EWSS, the twin-Fock state or the minimal quantum fluctuation state, which are commonly proposed for high precision measurements. 

Finally, we compare the numerical results for the operation times ${\tau}_{\mathrm{EWSS}}$, ${\tau}_{\mathrm{TFS}}$, 
and ${\tau}_{\Delta J_z}$, which are plotted in Fig~\ref{fig:tandmaxDz} (b) as functions of $J$. 
The evolution time ${\tau}_{\mathrm{EWSS}}$,  ${\tau}_{\mathrm{TFS}}$, and ${\tau}_{\Delta J_z}$ are respectively well fitted for all 
$J$ to 
\begin{equation} 
	{\tau}_{\mathrm{EWSS}} = \frac{\log {(1.10J)}}{4.02J}, \label{eq:FmaxtoEWSStime}
\end{equation} 
\begin{equation} 
	{\tau}_{\mathrm{TFS}} = \frac{\log {[(25.2\pm 0.2) J]}}{3.93J}, \label{eq:FmaxtoTFStime}
\end{equation} 
and
\begin{equation}
	\tau_{\Delta J_z} = \frac{\log {(11.5J)}}{3.94J}, \label{eq:DzmaxSSStime} 
\end{equation}
which implies that the SSS reaches the point where the minimum quantum fluctuation is achieved, i.e. $\langle (\Delta J_y)^2 \rangle$ 
becomes minimal, then the fidelity to the EWSS is maximized, and the fidelity to the twin-Fock state is maximized, 
and finally it reaches a state that gives the best sensitivity for the SSS. 

%\section{\label{sec:5}Discussions} 
To summarize, we have numerically analyzed the time evolution of SSSs under the two-axis counter twisting interaction and 
and the sensitivity in magnetic-field measurements. 
We find that at time ${\tau}_{\mathrm{EWSS}}$ the SSS can be approximately represented as the EWSS because of its 
high fidelity of $0.990$ in the large-$J$ limit, and after that, at the time ${\tau}_{\mathrm{TFS}}$ the SSS approximately becomes the 
twin-Fock state with the fidelity of $0.868$ in the large-$J$ limit.  
We also calculated the sensitivity defined by the lower bound of the Cram\'er-Rao inequality and show that 
the SSS reaches the Heisenberg limit and it exceeds the sensitivity limit given with the EWSS and the twin-Fock state, 
though it does not reach the sensitivity limit of the optimal state $|{\Psi}_{\mathrm{CAT}} \rangle$. 
To evaluate the feasibility, we still have to consider noise effects involved in the squeezing; however as there are theoretical proposals to realize two-axis counter twisting interaction in Bose-Einstein condensates and one-axis twisting has been already demonstrated beyond the standard quantum limit, we can expect that the SSSs could be more feasible than other states which realize the Heisenberg limit. 
The time requires to achieve the best sensitivity for the SSS is $\tau_{\Delta J_z} = \log {(11.5J)}/(3.94J)$, and we might be able to adjust the time and the size of the collective spin to minimize the effect of noise, though further research is necessary to clarify the noise effects.

\appendix 
\section*{ACKNOWLEDGMENTS} 
This work is supported by NICT(A), JSPS (Kiban-S), NTT, and Kakenhi Grant. No. 26287088. 

 
\end{document}